\documentclass[12pt]{article}

\usepackage{a4}
\begin{document}

\setcounter{page}{0}
\title{Lorentz Transformations from Reflections:\\ Some Applications}
\author{H.K. Urbantke\\
{\small Institut f\"ur Theoretische Physik,}
{\small Universit\"at Wien}\\
{\small Boltzmanngasse 5,}
{\small A-1090 Vienna, Austria}}
\date{3 December 2002}
\maketitle
\begin{abstract}

We point out, by exhibiting two examples and mentioning a third one, 
that it is sometimes useful to consider Lorentz transformations as 
generated from hyperplane or line reflections. One example concerns 
the construction of boosts linking two given 4-vectors, the other one 
concerns the Minkowski geometric understanding of V. Moretti's polar 
decomposition of orthochronous Lorentz matrices.

\end{abstract}
\newpage

\section{Introduction}

The Cartan-Dieudonn\'{e} theorem\footnote{See, e.g., [1]; Dieudonn\'{e} [2] 
contains a shortcut to the proof; the shortcut presented in [3] seems to 
be mistaken.}, to the effect that the orthogonal groups O($n$,C), O($p,q$) 
are generated by hyperplane reflections---each group element being a product 
of at most $n=p+q$ of them---is usually invoked only to study the structure 
of those groups in general and to establish their relation to Clifford 
algebras. However, the possibility of generating orthogonal transformations 
from reflections may sometimes be of some practical value in solving simple 
concrete problems involving orthogonal transformations. In this note we want 
to illustrate this point by discussing two problems involving Lorentz boosts  
from a geometrical rather than matrix calculational point of view. One 
concerns a geometric understanding of the construction of the boost matrix 
that carries two given non-spacelike future-directed 4-vectors into each 
other. The other concerns the geometrical understanding of V. Moretti's 
observation [4] that the usual rotation-boost decomposition (=Cartan 
decomposition, in the sense of the theory of symmetric spaces [5]) of an
orthochronous Lorentz matrix (cf.~[6] for an elementary approach) is a 
polar decomposition in the sense of (real) Hilbert space theory. In both 
cases we find that the true understanding of the elementary matrix 
solution comes from an analysis of boosts in terms of reflections. Another 
instance of the application of reflections to Lorentz transformations, 
using Clifford algebra in addition, is considered, e.g., in [7].

\section{Preparation: hyperplane and line reflections}

We work in a Lorentzian vector space {\bf V} of 4-vectors, i.e., a real
4-dimensional vector space with a symmetric bilinear form $\eta$ of 
signature $(-+++)$. Since we want to maintain a geometric language, it 
will be preferable to consider Lorentz transformations in their active 
interpretation as linear maps $L:{\bf V}\rightarrow{\bf V}$ satisfying 
$\eta(Lx,Ly)=\eta(x,y)$ for all $x,y\in{\bf V}$. By the Cartan-Dieudonn\'e 
theorem, each such $L$ is a product of at most four hyperplane reflections. 
A hyperplane reflection (more precisely, a reflection in a hyperplane) is 
a Lorentz transformation $S$ which is involutory (i.e., 
$S^2={\rm id}_{\bf V}\,\,\Rightarrow$ eigenvalues $\pm 1$) and such that the 
(+)eigenspace is a 3-plane and the (orthogonal) ($-$)eigenspace is 
1-dimensional and not lightlike. Thus, if $u$ with $\eta(u,u)\neq 0$ spans 
the ($-$)eigenspace, an arbitrary 4-vector $x$ decomposes as 
$x=x_{\perp}+x_{\parallel}$, where
$$x_{\perp}:=\frac{\eta(x,u)}{\eta(u,u)}u,$$
and the reflection $S_u$ in the hyperplane orthogonal to $u$ is given by 
$$x=x_{\parallel}+x_{\perp}\mapsto S_ux:=x_{\parallel}-x_{\perp}=x-2x_{\perp}.$$
Using (abstract) index notation, we find that $S_u$ is given by the tensor

$$\delta^{\mu}_{\nu}-2u^{\mu}u_{\nu}/\eta(u,u),$$
and the above-mentioned properties of $S_u$ are immediate from this 
construction. It is essential that $S_u$ is improper, $\det S_u=-1$.

The decomposition of a given $L$ as a product of hyperplane reflections is 
not unique; however, the parities of the numbers of reflections in 
timelike and in spacelike hyperplanes involved are unique and define the 
four components of the Lorentz group ${\cal L}$ as 
(even, even)$\Leftrightarrow{\cal L}^{\uparrow}_{+}$, 
(odd, even)$\Leftrightarrow{\cal L}^{\uparrow}_{-}$,
(even, odd)$\Leftrightarrow{\cal L}^{\downarrow}_{-}$,
(odd, odd)$\Leftrightarrow{\cal L}^{\downarrow}_{+}$ in the usual Wigner 
notation.

For the generation of the Lorentz group in 4 dimensions, hyperplane 
reflections may be replaced by line reflections, which are slightly easier 
to visualize. The reflection in the line spanned by the non-lightlike 
vector $u$ is just given by $-S_u$.
The essential feature of $S_u$ to be involutory and improper is preserved 
in this transition because of the assumed even dimensionality of {\bf V}.  
Using this we immediately see the following facts which we note down for 
future use.
1) We have $S_u x=-x$ or $S_u x=+x$ for $x\parallel u$ or $x\perp u$. 
2) If $u$ is timelike, $S_u$ is antichronous and is the time reversal 
with respect to an observer with 4-velocity $\parallel u$. 
3) If $x$ lies in the 2-plane spanned by two non-lightlike vectors $a$, 
$b$, then $S_a x$, $S_b x$, $S_a S_b x$ also lie in that plane. 
4) If $\eta(a,a)=\eta(b,b)$ (possibly =0) and $\eta(a+b,a+b)\neq 0$, then 
$-S_{a+b} b=a$ (reflection in the bisector of the Lorentzian angle between 
$a$ and $b$). (Of course, these statements can be checked by direct 
calculation as well.)

\section{Boosts and space rotations from reflections}

Lorentz boosts are characterized as proper orthochronous Lorentz 
transformations that leave invariant a timelike 2-plane as a whole and 
leave invariant vectorwise the orthogonal (spacelike) 2-plane. More 
specifically, a boost {\em relative to an observer} with 4-velocity 
$u$---or $u$-boost, for short---is 
a Lorentz transformation of this kind such that $u$ is contained in the 
invariant timelike 2-plane.

We shall now argue that all $u$-boosts can be written as $S_w 
S_u$, where $w$ is some other timelike 4-vector, determined up to 
proportionality and becoming unique by requiring it to be normalized and 
future-directed. Indeed, since $S_u$ and $S_w$ both are improper and 
antichronous, their product is proper and orthochronous; if $x$ is in the 
timelike 2-plane spanned by $u$ and $w$, so is $S_w S_u x$, and if $x$ is 
orthogonal to this plane, $S_u x=x$, $S_w x=x$, so $x$ is left fixed by 
$S_w S_u$: hence $S_w S_u$ is a $u$-boost. Conversely, if $B$ 
is a $u$-boost, carrying $u$ to $u':=Bu$, we can, as noted 
above, take $w$ to be a multiple of $u+u'$: then $S_w S_u u=S_w(-u)=u'=Bu$. 
So $S_{u'+u}S_u$ does the same job as $B$ and is therefore identical to $B$. 
Note that $u'+u$ is timelike and future-directed just as $u$, $u'$ are. 

Similarly, a proper space rotation {\em relative to an observer} 
with 4-velocity $u$---or $u$-rotation, for short---is a Lorentz 
transformation leaving invariant as a whole a spacelike 2-plane and leaving 
invariant vectorwise the orthogonal timelike 2-plane which must contain 
$u$. The direction in the latter orthogonal to $u$ gives the 
axis of rotation relative to $u$. We can generate such $u$-rotations 
simply as products $S_a S_b$, where $a$, $b$ are orthogonal to $u$ 
and span the invariant spacelike 2-plane. If they are normalized, 
$\eta(a,b)$ will give the cosine of half the rotation angle, as is well 
known from elementary geometry. Improper $u$-rotations are similarly generated 
by an odd number of hyperplane reflections with normals orthogonal to $u$. 

As a first application we mention the calculation of the {\em Thomas 
angle}: the product of a $u$-boost $B$ sending $u$ to $u'$, followed by a 
$u'$-boost $B'$ sending $u'$ to $u''$ and a $u''$-boost $B''$ sending 
$u''$ back to $u$ leaves $u$ invariant and is proper, so is a proper 
$u$-rotation; the angle of rotation involved is the Thomas angle. One can 
calculate it as $S_{u+u''}S_{u''}S_{u''+u'}S_{u'}S_{u'+u}S_u$, using 
Clifford multiplication for further evaluation and comparison with an 
expression $S_aS_b$ for $u$-rotations (see, e.g., [7]).

\section{Boosts linking two given 4-vectors}

In the preceding section, we constructed from reflections a $u$-boost 
that carries $u$ to another given 4-velocity $u'$. Here we 
want to find, more generally, a $u$-boost $B$ that carries a 
given 4-vector $x$ to another given 4-vector $x'$. Naturally, since $B$ is 
orthochronous, we must assume $x$, $x'$ not only to have the same 
non-positive 4-square $\eta(x,x)=\eta(x',x')\leq 0$ but also to have the 
same time orientation to make the problem well-posed.

It is, of course, just a clumsy exercise to find $B$ using the ordinary 
matrix form for boosts, i.e., to find the matrix $\underline{B}$ for $B$ 
in the rest frame of $u$, in particular to find the components {\bf v}
of the relative velocity involved in $\underline{B}$. Even more 
complicated is the algebra involved when the problem is related to  
relativistic velocity addition: a clear way of disentangling it has been 
developed only relatively recently (see [8] for exposition). However, 
using reflections we can give a surprisingly simple solution. Namely, as 
above we can write all $u$-boosts as $B=S_w S_u$, where $w$ is 
a timelike 4-vector to be determined. Requiring $x'=Bx=S_w S_u x$ we see 
that $w$ must be chosen such that $S_w$ carry $S_u x$ into $x'$. By the 
last remark of section 2, we see that under the assumptions made on 
$x$, $x'$ we have $w\propto x'-S_u x$, and the unique solution is

$$B=S_{x'-S_u x} S_u$$
whenever $x'-S_u x$ is not lightlike. The latter case arises only if $x'$ 
and $-S_u x$ are lightlike and parallel, i.e., if $x'$ and $x$ are 
lightlike and spatially antipodal with respect to $u$; in this case no 
$u$-boost exists that would do what is required. Otherwise, we 
see from the fact that $w$ bisects the Lorentzian angle between $u$ and 
$Bu=u'$ that the relative velocity {\bf v} involved in the matrix 
$\underline{B}$ is just the "relativistic double" of ${\bf w}/w^0$ in the 
rest frame of $u$. Thus, writing in this frame the column matrices of 
components 
\newline 
$\underline{x}=\left(\begin{array}{c}x^0\\{\bf x}\end{array}\right)$, 
$\underline{x'}=\left(\begin{array}{c}x'^0\\{\bf x}'\end{array}\right)$, 
we have 
$\underline{S_u x}=\left(\begin{array}{c}-x^0\\{\bf x}\end{array}\right)$, 
$\underline{w}=\left(\begin{array}{c}x'^0+x^0\\{\bf x}'-{\bf x}
\end{array}\right)$, 
whence 

$${\bf v}=\frac{2(x'^0+x^0)({\bf x}'-{\bf x})}{(x'^0+x^0)^2+({\bf x}'-{\bf 
x})^2},$$
where the speed of light has been set to $c=1$.

We remark that the occurrence of the exceptional case above is a symptom 
of the following fact which plays a role in the massless helicity 
representations of the Poincar\'{e} group: if $x$, $x'$ are lightlike
and future-orientd, one cannot find Lorentz transformations carrying a 
fixed $x$ to a variable $x'$ such that the transformation depends on $x'$ 
in a fashion {\em continuous on the entire future light cone} (see [6], 
[9] for discussion).

\section{Symmetry and positivity for boosts}

Recently, Moretti [4] observed that the usual polar decomposition of real 
matrices when applied to orthochronous Lorentz matrices gives just their 
rotation-boost decomposition. Indeed, as remarked in [10], this follows 
already from the uniqueness of the polar decomposition, since Lorentz 
matrices describing (proper or improper) spatial rotations are trivially 
orthogonal (in the 4-dimensional Euclidean sense!) and boost matrices 
$\underline{B}$ are symmetric and positive definite (in the same sense), 
the latter because of the identity

$$(t\,\,{\bf x}^{\top})\underline{B}\left(\begin{array}{c}t\\{\bf 
x}\end{array}\right)
\equiv \gamma(t-{\bf vx})^2+\frac{1}{\gamma}\frac{({\bf vx})^2}{{\bf v}^2}
+({\bf x}-\frac{{\bf vx}}{{\bf v}^2}{\bf v})^2.$$
(For the active interpretation, the sign of the boost velocity {\bf v} 
must be reversed.)

We here want to provide a geometrical understanding of the latter 
properties, which are not immediate in a Lorentzian context, because the 
only geometric structure around in {\bf V} seems to be $\eta$ (and space+time 
orientation). What then is the geometric meaning of the "4-dimensional 
Euclidean sense" here, and how does the identity above look geometrically? 
It turns out that again reflections are helpful in clearing this up.

The "hidden" unit matrix defining the Hilbert space scalar product to 
which the terms "symmetric" and "positive-definite" refer in the polar 
decomposition context is a basis-dependent 
object, and the basis in {\bf V} to which it refers contains the observer 
4-velocity $u$ as one of its constituents. This remark allows an 
{\em observer-dependent} Hilbert space scalar product to be 
geometrically defined as $U(x,y)=U_{\mu\nu}x^{\mu}y^{\nu}$, 
where the tensor $U$ is given by 

$$U_{\mu\nu}:=\eta_{\mu\nu}-2u_{\mu}u_{\nu}/\eta(u,u):$$ 
in the rest frame of $u$ its components constitute the unit matrix.
Then what we want to see is that every $u$-boost $B$ satisfies

$$U(x,By)=U(y,Bx)\,\,\,{\rm for\, all}\,\,x,y\in {\bf V}$$
$$U(x,Bx)> 0\,\,\,{\rm for\, all}\,\,x\neq 0.$$

We invoke the possibility of writing $B=S_w S_u$ for some timelike 
4-vector $w$ which we may also assume normalized and future-directed. 
Interpreting $\eta$, $U$ as maps from {\bf V} to its dual and looking at 
the explicit form for $S_u$ written before we obviously have $U=\eta S_u$; 
and similarly we have that $W:=\eta S_w$ relates $w$ and $S_w$ to a symmetric 
positive-definite bilinear form $W$. Using $S_u^{\top}\eta S_u=\eta$, 
$S_u^2={\rm id}_{\bf V}$ (which says that $S_u$ is an 
involutory Lorentz transformation) we now calculate

$$UB=\eta S_u S_w S_u=(S_u^{\top})^{-1} \eta S_w S_u=S_u^{\top} W S_u,$$
implying 

$$U(x,By)=W(S_ux,S_uy)$$ 
for all $x$, $y$. The stated properties of $W$ thus imply the properties 
in question.

For $u$-rotations, proper or improper, $U$-orthogonality follows from 
$S_a^{\top} U S_a=U$ whenever $\eta(a,u)=0$; but this is less surprising.

\section{Conclusion}

We mentioned one and gave two more applications of the fact that Lorentz 
transformations, and boosts in particular, can be generated from 
reflections. There are probably many more instances where this point of 
view might be useful in clearing up the geometrical structure and/or 
replacing heavier matrix calculations. Our aim was to bring this point to 
fore.

\section*{References}

\begin{description}

\item[[1]] \'{E}. Cartan, {\em The Theory of Spinors}. MIT Press, 
Cambridge (Mass.) 1966.

\item[[2]] J. Dieudonn\'{e}, {\em Sur les groupes classiques}. Paris, 
Hermann 1948. 

\item[[3]] W. Greub, {\em Multilinear Algebra}, $2^{nd}$ edition. 
Springer, NewYork 1978.

\item[[4]] V. Moretti, The Interplay of the Polar Decomposition Theorem 
and the Lorentz Group. arXiv:math-ph/0211047.

\item[[5]] J. Dieudonn\'{e}, {\em Treatise on Analysis}, vol.~5/6. 
Academic Press, NewYork 1977.

\item[[6]] R.U. Sexl, H.K. Urbantke, {\em Relativity, Groups, Particles. 
Special Relativity and Relativistic Symmetry in Field and Particle 
Physics}. Springer, Wien NewYork 2001.

\item[[7]] H.K. Urbantke, Physical Holonomy, Thomas Precession, and 
Clifford Algebra. Am. J. Phys. {\bf 58}, 747 (1990); {\bf 59}, 1150 
(1991).

\item[[8]] A.A. Ungar,{\em Beyond the Einstein Addition Law and its 
Gyroscopic Thomas Precession. The Theory of Gyrogroups and Gyrovector 
Spaces}. Kluwer, Dordrecht 2001.

\item[[9]] H.K. Urbantke, The Hopf Fibration---Seven Times in Physics. J. 
Phys. Geom. 2001 (in print).

\item[[10]] H.K. Urbantke, Elementary proof of Moretti's polar 
decomposition theorem for Lorentz transformations. arXiv:math-ph/0211077.

\end{description}

\end{document}